\documentclass[runningheads]{llncs}
\usepackage[T1]{fontenc}
\usepackage{graphicx}
\usepackage{multirow}
\usepackage{hyperref}
\usepackage{array}
\usepackage{cite}
\usepackage{enumitem}

\newcolumntype{L}[1]{>{\raggedright\let\newline\\\arraybackslash\hspace{0pt}}m{#1}}
\newcolumntype{C}[1]{>{\centering\let\newline\\\arraybackslash\hspace{0pt}}m{#1}}
\newcolumntype{R}[1]{>{\raggedleft\let\newline\\\arraybackslash\hspace{0pt}}m{#1}}
\usepackage{color}

\newcommand{\dq}[1]{\href{https://stackoverflow.com/questions/#1/}{$A_{#1}$}}
\newcommand{\ques}[1]{\textit{`#1'}}
\begin{document}
\title{Can  Transformer  Models  Effectively  Detect Software  Aspects  in  StackOverflow  Discussion?}
\titlerunning{Performance of Transformers in Software Aspects Detection}
%
\author{Nibir Chandra Mandal \and
Tashreef Muhammad \and
G. M. Shahariar}
\authorrunning{Mandal et al.}
%
\institute{Ahsanullah University of Science and Technology, Dhaka, Bangladesh \\
\email{\{nibir338,tashreef.muhammad,sshibli745\}@gmail.com}}

\maketitle              
\begin{abstract}
Dozens of new tools and technologies are being incorporated to help developers, which is becoming a source of consternation as they struggle to choose one over the others. For example, there are at least ten frameworks available to developers for developing web applications, posing a conundrum in selecting the best one that meets their needs. As a result, developers are continuously searching for all of the benefits and drawbacks of each API, framework, tool, and so on. One of the typical approaches is to examine all of the features through official documentation and discussion. This approach is time-consuming, often makes it difficult to determine which aspects are the most important to a particular developer and whether a particular aspect is important to the community at large. In this paper, we have used a benchmark API aspects dataset (Opiner) collected from StackOverflow posts and observed how Transformer models (BERT, RoBERTa, DistilBERT, and XLNet) perform in detecting software aspects in textual developer discussion with respect to the baseline Support Vector Machine (SVM) model. Through extensive experimentation, we have found that transformer models improve the performance of baseline SVM for most of the aspects, i.e., `Performance', `Security', `Usability', `Documentation', `Bug', `Legal', `OnlySentiment', and `Others'. However, the models fail to apprehend some of the aspects (e.g., `Community' and `Potability') and their performance varies depending on the aspects. Also, larger architectures like XLNet are ineffective in interpreting software aspects compared to smaller architectures like DistilBERT.

\end{abstract}

\section{Introduction}
\label{sec:introduction}
The information technology (IT) sector is leading the fourth industrial revolution and bringing about significant developments in the $21^{st}$ century. The advancement of the sophisticated field is accelerating thanks to the concept of ``open source''. By deploying developed modules from various developers under a Creative Commons (CC) License \cite{cc-intro} and the like, the opportunity to develop more and more over existing technologies has become quite simple. APIs are a popular technology these days. APIs enable third-party applications to easily access information while maintaining data security.
APIs can now be found everywhere due to the widespread availability of various resources. It is challenging to identify the suitable API to get the task done. The ``Opiner'' \cite{gias_summary} was proposed as a solution to this challenging problem. Opiner is a website that provides multiple summarized thoughts about APIs in an online search engine. Opiner's backend is powered by ``Automatic Summarization of API Reviews'' that employs the pattern recognition model Support Vector Machine (SVM) \cite{uddin2017automatic}. The model was trained using around 4000 sentences from \textit{Stack Overflow} \footnote{https://stackoverflow.com/}.
The advent of \textit{Stack Overflow} has been a blessing to developers, as it has increased the frequency with which millions of engineers from various communities discuss the same subjects. This allows the newcomer's most welcoming environment to post their relatively unsophisticated difficulties. Furthermore, the cross-domain conversation is available here. "What are the better alternatives to pandas?" for example. One of the typical
approaches is to examine all of the features through official documentation and discussion, which is time-consuming and can make it difficult to determine which aspects are the most important to a particular developer. Furthermore, most of the time, such conversations are ignored by the community. In addition, it is difficult to determine whether a
particular aspect is important to the community at large.  Despite these options, developers still find it difficult to choose the best solutions. This is due to the vastness and scarcity of posts. Each post may have several aspects, such as security and performance, or none at all. As a result, it is required to separate the portion that contains any aspects as well as the aspects linked with those. Thus, in this paper, we have concentrated on detecting aspects in textual developer discussion. As posts may contain multiple aspects, we analyze sentence level aspect detection. In summary, we make the following contributions in this paper: 
\begin{itemize}
    \item To increase the performance of the baseline Support Vector Machine (SVM) based model, we used four other Transformer models (BERT, RoBERTa, DistilBERT, and XLNet) and observed considerable performance improvement in most situations. 
    \item We conducted a thorough performance comparison and discovered that transformer models completely fail to understand some aspects, and their performance varies depending on the aspects, through rigorous experiments. In addition, as compared to smaller designs like DistilBERT, bigger architectures like XLNet are inadequate at analyzing software aspects.
\end{itemize}

\section{Related Work}
\label{sec:related_work}

\subsection{Literature Review}
Several recent papers have used SO posts to investigate various aspects of software development using topic modeling, such as what developers are talking about in general \cite{barua}. Mandal et al. \cite{gias_emp} investigated at nearly 53,000 IoT-related posts on SO and used topic modeling \cite{blei} to figure out what people were talking about. They intended to learn about the practical difficulties that developers experience when building real IoT systems. Recent studies \cite{kavaler, parnin} explored the connection between API usage and Stack Overflow discussions. Both research discovered a relationship between API class use and the number of Stack Overflow questions answered. But Gias Uddin \cite{gias_auto} utilized their constructed benchmark dataset named "OPINER" \cite{gias_mining} to carry out the study and noticed that developers frequently provided opinions about vastly different API aspects in those discussions which was the first step towards filling the gap of investigating the susceptibility and influence of sentiments and API aspects in the API reviews of online forum discussions.
Uddin and Khomh \cite{gias_mining} introduced OPINER, a method for mining API-related opinions and providing users with a rapid summary of the benefits and drawbacks of APIs when deciding which API to employ to implement a certain feature. Uddin and Khomh \cite{gias_summary} used an SVM-based aspect classifier and a modified Sentiment Orientation algorithm \cite{liu} to comply with API opinion mining. Based on the positive and negative results emphasized in earlier attempts to automatically mine API opinions, as well as the seminal work in this field by Uddin and Khomh \cite{gias_summary}, Lin et al. \cite{lin_bin} introduced a new approach called Pattern-based Opinion MinEr (POME), which utilizes linguistic patterns preserved in Stack Overflow sentences referring to APIs to classify whether a sentence is referring to a specific API aspect (functional, community, performance, usability, documentation, compatibility or reliability), and has a positive or negative polarity.
Some other research works have concentrated on mining views in order to acquire knowledge about API usage. Wang et al. \cite{shaohua} mined Stack Overflow for brief practical and beneficial API use advice from developer replies. DEEPTIP, their suggested method, used Convolutional Neural Network (CNN) with a dataset of annotated texts (classified as "tip" or "non-tip") and achieved a high precision of 80\%. Zhang and Hou \cite{zhang} extracted Oracle's Java Swing Forum online conversations for problematic API features. The HAYSTACK technique they developed recognized negative statements using a sentiment analysis approach and analyzed these unfavorable comments using pre-defined grammatical patterns to reveal problematic characteristics.

\subsection{Comparison}
The main difference between the study of this paper to the existing studies is that the paper introduces usage of transformer type models. The study revolts around NLP and transformers are known for performing well in the sector of NLP. Hence, the main contribution of the study is introducing transformer which has not been done before.

\section{Background Studies}
\label{sec:BStudy}
\subsection{StackOverflow}
\label{sec3:stackoverflow}
\begin{figure}[h]
    \centering
    \includegraphics[width = 0.95\textwidth, keepaspectratio]{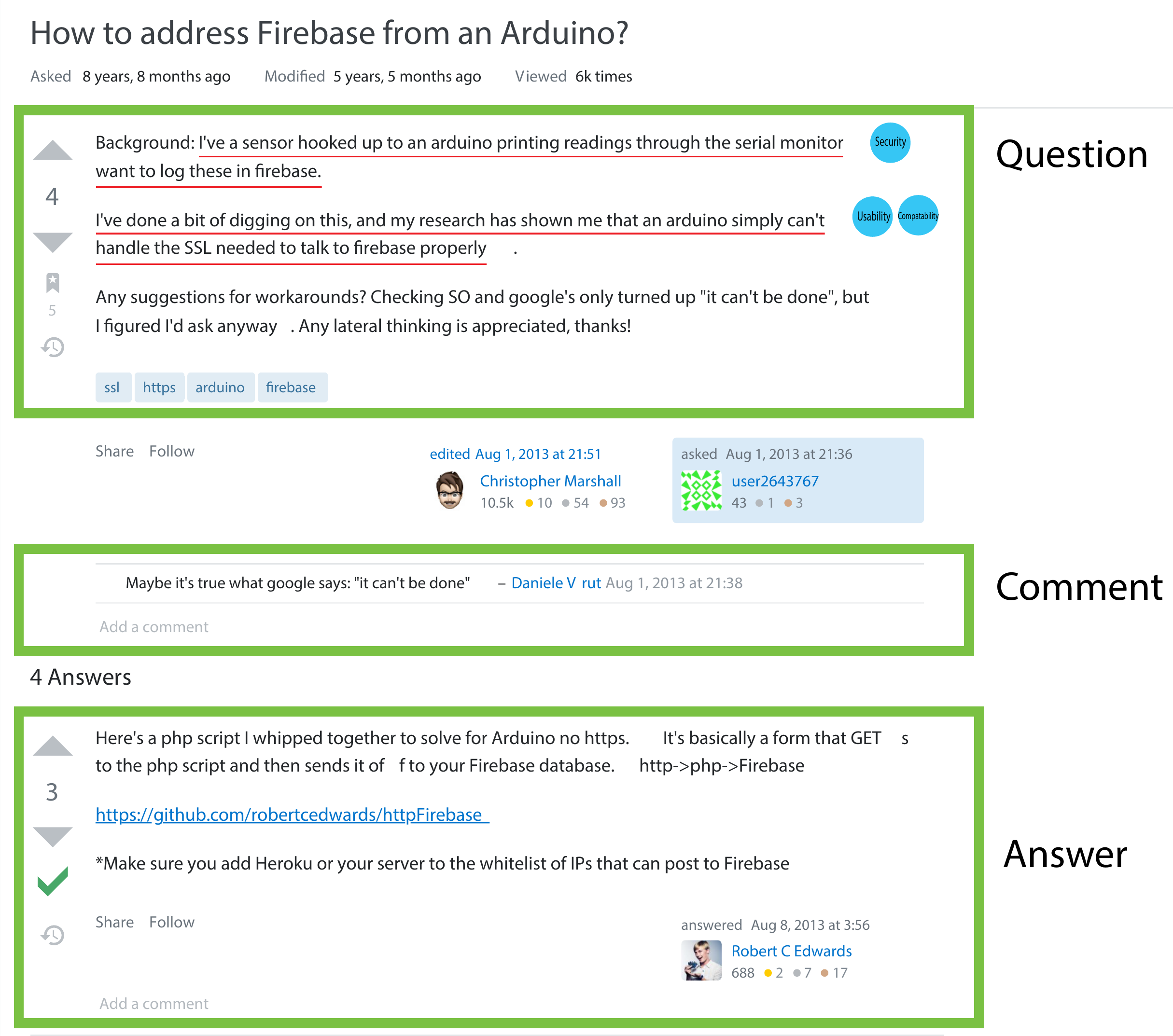}
    \caption{Different Parts of a Stack Overflow Post}
    \label{fig:structure_stackoverflow}
\end{figure}

Stack Overflow (SO) is a question and answer website for programmers \cite{stackoverflow-wiki}. It is an open platform where many people ask questions and give answers related to programming. A typical Stack Overflow post contains a question, some number of answers, and some comments as shown in Figure ~\ref{fig:structure_stackoverflow}. 

\subsection{Software API}
\label{sec3:software_api}
Application Programming Interface (API) is one of the most commonly used tools while developing software based services and products in recent times. It offers an easy accessibility (sometimes with enhanced security) of data very easily. Software developers use software APIs with appropriate category and architecture to get their job done easily. Some of the most popular APIs in RapidAPI include: \textit{Skyscanner Flight Search}, \textit{Open Weather Map}, \textit{API-FOOTBALL}, \textit{The Cocktail DB} and such \cite{example-api}.

\subsection{Software API Aspects}
\label{sec3:software_api_aspects}
An API may have multiple aspects associated with it, and developers choose their desired API based on those aspects and comparing them with their necessities.
Following the survey and studies conducted by Uddin et al. \cite{uddin2019understanding}, the study proposed a total of 11 aspects that are used for assessing a software API by coders or developers. Those aspects are as follows:

\begin{enumerate}[label=\textbf{(\alph{*})}, wide, labelwidth=!, labelindent=0pt]
    \item \textbf{Performance:}
            This aspect includes comparison between two or more APIs in terms of speed, resource usages, etc. For example, a developer comments \textit{`The alternative is to poll for changes at regular intervals, but that may create a delay in the event reaching a client.'} in SO question \dq{1728020}. This sentence are related to the performance aspect of `Java HTTP'. 
    \item \textbf{Usability:}
            This aspect discusses the usages of an API, the difficulties of integrating an API, etc. are related to this aspect. For example, SO question,\dq{1728020}, has a usability aspect related sentence as follows: \ques{All of this assumes we have a client that the server needs to be able to push data to.} 
    \item \textbf{Security:}
            This aspect is about how much security of data is ensured in the API. For instance, SO questions \dq{992019} contains the following password encryption related security discussions: \ques{I have tried a 32bit one for 256bit encryption, but it did not work as expected.}
    \item \textbf{Documentation:}
            This aspect is about how clear and thorough the official documentation is of the API. To give an example, the SO question \dq{2971315} can be considered where one of the answers \ques{From the API documentation: As of release JDK 5, this class has been supplemented with an equivalent class designed for use by a single thread, \textbf{StringBuilder}.} is related to the documentation relating to 'StringBuilder'.
    \item \textbf{Compatibility:}
            This aspect is about how easily the API is compatible to the given framework environments. In the SO question \dq{5059224}, the question regarding \textit{Which is the best library for XML parsing in java} was answered by someone with \ques{JDOM would be another alternative to DOM4J} is an example of compatibility for Java based XML parsing.
    \item \textbf{Portability:}
            This aspect is about how portable is the API in situations like being used in multiple Operating System environments. The SO question in \dq{1109307} has an answer saying \ques{These are standard on Linux, are generally available on Unix and can be run on Windows under cygwin, although you may be able to find windows-native apps that can do it as well} is an example of portability.
    \item \textbf{Community:}
            This aspect is about how much active is the community of users who use the API for practice. In the SO question \dq{326390}, the discussion contains \ques{And it seems to be the most wide-spread, at least in the sites I've visited} is an example of community aspect.
    \item \textbf{Legal:}
            This aspect is about how much proper licensing and access is provided for using the API. In the SO question \dq{2129375} which is related to 'HTML/XML Parser for Java' an answer quotes the use of 'Apache Tika' as the best choice. One of the reasons behind his explanation includes a sentence \ques{Furthermore, it is open-source} which directly indicates towards the legal aspects of the API that is being discussed.
    \item \textbf{Bug:}
            This aspect is about the presence or absence of bug or errors in the API overall. In the SO question \dq{507391}, there is a discussion regarding comparison between 'Active MQ JBOSS Messaging' where the answer \ques{Feature-by-feature comparisons are all very well, but my experience of ActiveMQ (through various versions over the years) is that it is shockingly buggy, and no one seems inclined to fix those bugs} clearly indicates the buggy features of 'Active MQ' that will make anyone think twice about using the API. This is an example of the bug aspect of an API.
    \item \textbf{Only Sentiment:}
            This aspect expresses only sentiment about an API without technical brief on the matter. The simple sentence of \ques{It's deeply frustrating} in the SO question \dq{507391} is a prime example of using sentiment only to express the aspect of an API.
    \item \textbf{Others:}
            The aspect that does not fall under all the other mentioned categories are stated in here. Even after specifying 10 concrete aspects, there are far too many types of sentences that are found in real-life relating to the performance of APIs. All those are gathered in this category. For example, the sentence \ques{I can manually change it, but I'd rather not have to remember to update it every time the jaxb files are regenerated} can be considered as an example of such aspects. This example was taken from SO question \dq{15936368}.
\end{enumerate}
\noindent Developers primarily follow these aspects before selecting an API for their projects. They analyze the sentiment of these aspects and make a decision accordingly. 
\subsection{Transformers}
A neural network that learns factors in sequential data is referred to as a transformer model. In natural language processing transformer models are used frequently as it tracks the meaning of consecutive data by tracking the relationships among them like the words in a sentence. Transformer models were first introduced by Google \cite{transformer} and are one of the most powerful models till date in the history of computer science.
We incorporated four different kinds of transformer model in our work which are RoBERTa, BERT, DistilBERT and XLNet.  Bidirectional Encoder Representations from Transformers or BERT \cite{bert} was first established by Jacob Devlin et al. which is a transformer model to represent language. Yinhan Liu et al. introduced RoBERTa \cite{roberta} as a replication study of BERT \cite{bert}. Authors showed that their model was highly trained than BERT overcoming the limitations and showed a good performance over basic BERT model. 
DistilBERT was proposed by Victor Sanh et al. \cite{distilbert} as a general purpose transformer model which is quite smaller than other language representation models. Authors showed that the size of a BERT model can be lessened by $40\%$ by leveraging knowledge filtering at the time of pre-training stage. They also showed that their model is $60\%$ faster than other transformer models. Zhilin Yang et al. introduced XLNet \cite{xlnet} which overcomes the constraints of BERT \cite{bert} using a universal auto-regressive pre-training technique. Maximum expected likelihood was taken into consideration over all arrangements of the factorization order. They have showed that their model beat BERT by a huge margin on a variety of tasks like sentiment analysis, language inference or document ranking etc.
All these used models, RoBERTa, BERT, XLNet and DistilBERT have different architectures. The Table ~\ref{tab:transformer_parameter} contains details about these architectures.

\begin{table}[htbp]
    \centering
    \caption{Architecture details of used Transformer variants}
    \label{tab:transformer_parameter}
    \begin{tabular}{l|crrrr} \hline
    \textbf{Architecture} & \textbf{Used Model} & \textbf{Layers} & \textbf{Heads} & \textbf{Hidden} & \textbf{Parameters} \\ \hline 
    RoBERTa &  distilroberta-base & 6 & 768 & 12 & 82M \\
    BERT & bert-base-uncased & 12 & 768 & 12 & 110M \\
    XLNet & xlnet-base-cased & 12 & 768 & 12 & 110M \\
    DistilBERT & distilbert-base-uncased & 6 & 768 & 12 & 66M \\ \hline
    \end{tabular}
\end{table}

\section{Dataset}
\label{sec:dataset}

Previously, Uddin et al.~\cite{gias_mining} studied API aspects in SO. We follow the study and use benchmark dataset. The dataset was collected from SO posts, comments, and answers. It was primarily created for mining developers opinions. Thus, it includes varieties of opinions about multiple APIs. It contains 4522 sentences and each sentences are associated with one or more aspects. 
To keep the context of the dataset more concentrated on the textual data itself, some slight modifications were added. Initially the dataset consisted of some threads. Each thread was either a question, a comment or an answer of a SO post. It contained data wrapped with some HTML tags. These data were formatted for the above mentioned purpose of making it more concentrated on the textual format. For example,
\begin{itemize}
    \item All hyperlinks were formatted by removing the HTML formats and appending a \textit{URL\_} in the prefix.
    \item Code examples were removed and replaced with \textit{CODESNIPPET} and \textit{CODETERM} as placeholders.
\end{itemize}
For example, the following sentences can be considered. \textbf{\textit{Example 1}} contains two aspects, i.e., \textit{Bug} and \textit{Performance}. Again, sentences like in \textbf{\textit{Example 2}} contains \textit{Only Sentiment} aspect. Also, in Figure ~\ref{fig:structure_stackoverflow} there are two underlined sentences. One sentence only contains the \textit{Security} aspects whereas the other sentence contains \textit{Usability} and \textit{Compatibility} aspects. \\

\begin{minipage}{0.95\textwidth}
    \label{exmp1}
    \textit{\textbf{Example 1:} What HTML parsers have the following features: Fast Thread-safe Reliable and bug-free Parses HTML and XML Handles erroneous HTML Has a DOM implementation Supports HTML4, JavaScript, and CSS tags Relatively simple, object-oriented API What parser you think is better?}
\end{minipage} \\

\begin{minipage}{0.95\textwidth}
    \label{exmp2}
    \textit{\textbf{Example 2} This works fine for the following code: CODESNIPPET\_JAVA2}
\end{minipage} \\

\noindent Table \ref{tab:dataset_distribution} shows the summary of our benchmark dataset. The dataset are clearly imbalanced and most of aspects have a positive rate below 5\%. Others aspect has the highest number of samples (i.g., 1699 samples ) followed by Usability aspect (1437 samples), Performance aspect (348 samples), Only sentiment aspect (348 samples), Documentation aspect (253 samples), Security aspect (253 samples), and so on. Legal aspect has the lowest number of samples- i.e., only 50 samples. This distribution of topics in our benchmark dataset indicates that developers most frequently discuss about Usability aspect in SO and they rarely discuss about Legal aspects of APIs. As usability, security, and performance are the most widely used aspects of APIs in developer community, we find enough discussion about those in our benchmark dataset.

\begin{table}[htbp]
    \centering
    \caption{Aspects distribution in the benchmark dataset}
    \label{tab:dataset_distribution}
    \begin{tabular}{lr||lr} \hline
        \multicolumn{1}{c}{\textbf{Aspects}} & \textbf{\# of Samples}    &   \multicolumn{1}{c}{\textbf{Aspects}} & \textbf{\# of Samples} \\ \hline \hline
        Performance	    &   348 (8\%)   &   Usability	    &   1437 (32\%) \\
        Security	    &   163	  (4\%) &   Documentation	&   253	  (6\%) \\
        Compatibility	&   93	  (2\%) &   Portability	    &   70	  (2\%) \\
        Community	    &   93	  (2\%) &   Legal	        &   50	  (1\%) \\
        Bug	            &   189	  (4\%) &   Only Sentiment  &   348	  (8\%) \\
        Others	        &   1699  (38\%) &  &   \\ \hline
    \end{tabular}
\end{table}
\vspace{-2em}
\section{Methodology}
\label{sec:proposed_methodology}
The proposed system for aspect categorization is presented in this section. In this study, whole experiment setup is divided into five major steps. These key phases or steps of the proposed method for binary aspect classification (\textit{considering Usability aspect as an example}) are summarized in Figure~\ref{classifier} and are further detailed below.
\begin{enumerate}[label=\textbf{Step \arabic{enumi})}, wide, labelwidth=!, labelindent=0pt]
\item \textbf{Input Sentence}: Each raw sentence from the dataset is presented to the proposed model one by one for further processing. Before that each sentence goes through some pre-processing steps. For example, all urls and codes are removed and for some special entries are replaced by specific placeholders. These information actually do not contain any validation to the aspect considering the variation it creates in the sample inputs. Hence, they are summarized to reduce load on the model.
\item \textbf{Tokenization}: Once the input sentence is gathered, each processed sentence is tokenized using the \textit{BERT Tokenizer} \cite{bert}. This is a well known tokenizer that applies an end-to-end, text string to wordpiece tokenization. The system utilizes wordpiece tokenization after applying basic tokenization. Each tokenized sentence gets a length of 100 tokens and zero padded when required. In the event of length more than 100, we cut off after 100. In other words, for length greater than 100, it gets truncated to keep everything in synchronization. The output of this step is a tokenized sentence of size 100.
\item \textbf{Embedding}: The tokenization step converts a complete sentence into tokenized list of sentences. However, it is not yet mathematically possible to compute it. In this regard, word embedding needs to be utilized. For word embedding, BERT \cite{bert} is used to turn each token in a sentence into a numeric value representation. Each token is embedded by $768$ real values via BERT. The input to this step is a tokenized sentence of size 100 and output of this step is an embedded sentence of size $100 \times 768$. Now each of the sentences that were represented by list of tokens containing the length of 100 is now converted a 2D numeric list with each component of the list expressed by an array of 768 numbers.
\item \textbf{Pooling}: To reduce the dimension of the feature map ($100 \times 768$) for each tokenized sentence in step 3, max pooling is used. It provides a real valued vector representation of size 768 per sentence. The resultant of this step is  a significant reduction of size (by a hundred) and yet keeping the integrity of the data intact.
\begin{figure}
\includegraphics[width=\textwidth]{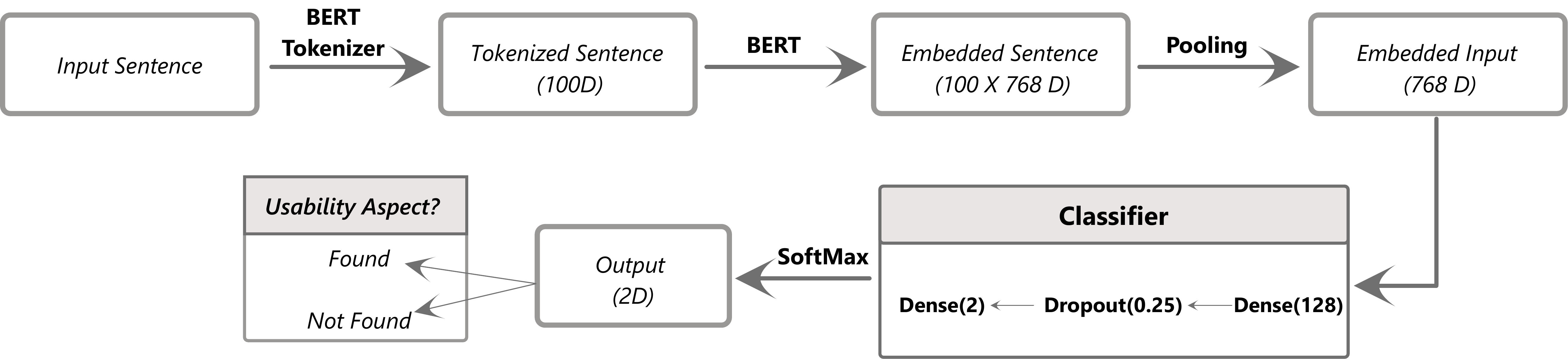}
\caption{Aspects Classification Process} \label{classifier}
\end{figure}
\item \textbf{Classification:} The data is currently ready for entering the training phase. Thus the current step is to prepare the classification model. Aspect classification is accomplished by the use of transfer learning \cite{transferlearning}. In total the goal is to classify whether a sentence belongs to one of 11 categories or aspects or not. To detect 11 various aspects, a basic neural network with two dense layers (with 128 and 2 neurons) is utilized to fine-tune the pretrained four different transformer models one by one such as RoBERTa \cite{roberta}, BERT \cite{bert}, DistilBERT \cite{distilbert}, and XLNet \cite{xlnet} without freezing any of the previous layers of the existent pre\-trained model. The pre\-trained weights were initialized using \textit{glorot\_uniform} in the simple neural network. A dropout rate of 0.25 is employed between the dense layers. Throughout the scope of this experiment, there were a total of four pre\-trained models that were fine-tuned to serve the purpose of classification. 
\item \textbf{Aspect detection:} Because a sentence may have several aspects, we have employed 10-fold cross validation and binary classification. This helps evaluating the performance of the model more rigorously. On the output layer, the \textit{Softmax} activation function is used. The final result of the model is to detect whether or not a specific aspect has been found on the provided input sentence.
\end{enumerate}

\section{Result Analysis}
\label{sec:result_analysis}
\subsection{Experimental Results}\label{sec: experimental_result}

\begin{table}[h]
    \centering
    \caption{Performance in different models}
    \label{tab:perf_list}
    
    \resizebox{\textwidth}{!}{\begin{tabular}{|c|c||C{6em}|C{6em}|C{6em}|C{6em}||C{6em}|}
    \hline
    \multirow{2}{*}{\textbf{Aspect}} & \multirow{2}{*}{\textbf{Metric}} &  \multicolumn{4}{c||}{\textbf{Transformer Models}} & \textbf{Baseline} \\ \cline{3-6}
     &   &   \textbf{RoBERTa}  &   \textbf{BERT}   &   \textbf{XLNet}  & \textbf{DistilBERT}   & \textbf{SVM} \\ \hline \hline
    \multirow{3}{8em}{Performance} & P & \textbf{0.80} & 0.76 & 0.78 & 0.78 & 0.78 \\
     & R & 0.74 & \textbf{0.77} & 0.71 & 0.71 & 0.46 \\
     & F1 & \textbf{0.77} & 0.76 & 0.75 & 0.74 & 0.56 \\ \hline \hline
    \multirow{3}{8em}{Usability} & P & \textbf{0.68} & 0.67 & 0.61 & 0.66 & 0.53 \\
     & R & \textbf{0.69} & 0.67 & 0.65 & 0.67 & 0.75 \\
     & F1 & \textbf{0.68} & 0.67 & 0.63 & 0.66 & 0.62 \\ \hline \hline
    \multirow{3}{8em}{Security} & P & \textbf{0.81} & 0.72 & 0.65 & 0.81 & 0.78 \\
     & R & 0.74 & \textbf{0.81} & 0.73 & 0.74 & 0.58 \\
     & F1 & 0.72 & 0.75 & 0.68 & \textbf{0.76} & 0.60 \\ \hline \hline
    \multirow{3}{8em}{Community} & P & 0.18 & 0.19 & 0.22 & \textbf{0.33} & 0.40\\
     & R & 0.13 & 0.11 & \textbf{0.17} & 0.16 & 0.24 \\
     & F1 & 0.15 & 0.13 & 0.15 & \textbf{0.20} & 0.26\\ \hline \hline
    \multirow{3}{8em}{Compatibility} & P & 0.17 & 0.10 & 0.10 & \textbf{0.30} & 0.50 \\
     & R & 0.04 & 0.01 & 0.01 & \textbf{0.09} & 0.08 \\
     & F1 & 0.06 & 0.02 & 0.02 & \textbf{0.13} & 0.13 \\ \hline \hline
    \multirow{3}{8em}{Portability} & P & \textbf{0.67} & 0.61 & 0.33 & 0.63 & 0.63 \\
     & R & 0.45 & 0.49 & 0.15 & \textbf{0.62} & 0.63 \\
     & F1 & 0.49 & 0.51 & 0.19 & \textbf{0.59} & 0.61 \\ \hline \hline
    \multirow{3}{8em}{Documentation} & P & \textbf{0.69} & 0.69 & 0.59 & 0.65 & 0.59 \\
     & R & \textbf{0.57} & 0.50 & 0.50 & 0.54 & 0.43 \\
     & F1 & \textbf{0.61} & 0.56 & 0.53 & 0.57 & 0.49 \\ \hline \hline
    \multirow{3}{8em}{Bug} & P & 0.64 & \textbf{0.70} & 0.59 & 0.65 & 0.57 \\
     & R & 0.63 & 0.63 & 0.58 & \textbf{0.66} & 0.50 \\
     & F1 & 0.61 & \textbf{0.63} & 0.56 & 0.63 & 0.51 \\ \hline \hline
    \multirow{3}{8em}{Legal} & P & \textbf{0.72} & 0.44 & 0.31 & 0.71 & 0.70 \\
     & R & \textbf{0.62} & 0.39 & 0.42 & 0.61 & 0.46 \\
     & F1 & 0.61 & 0.38 & 0.35 & \textbf{0.63} & 0.52 \\ \hline \hline
    \multirow{3}{8em}{OnlySentiment} & P & \textbf{0.69} & 0.69 & 0.69 & 0.65 & 0.61 \\
     & R & \textbf{0.51} & 0.48 & 0.50 & 0.48 & 0.43 \\
     & F1 & \textbf{0.58} & 0.54 & 0.57 & 0.55 & 0.50 \\ \hline \hline
    \multirow{3}{8em}{Others} & P & 0.77 & 0.74 & \textbf{0.77} & 0.74 & 0.61 \\
     & R & \textbf{0.71} & 0.68 & 0.64 & 0.68 & 0.67 \\
     & F1 & \textbf{0.73} & 0.70 & 0.69 & 0.71 & 0.64 \\ \hline
    \end{tabular}}
\end{table}

We show our experimental results in Table \ref{tab:perf_best}. For the most part, we discovered that transformer models outperform baseline SVM. However, for Community and Portability aspects, the SVM model still holds the highest performance. In addition, we found that the performance of each transformer varies across the different aspects. The results for each aspect are as follows:

\begin{enumerate}[label=\textbf{(\alph{*})}, wide, labelwidth=!, labelindent=0pt]

\item \textbf{\textit{Performance }} We found that RoBERTa offers the best results in terms of Precision and F1-Score, whereas BERT shows the best recall. In addition, RoBERTa's F1-Score of 0.77 beats baseline SVM's 0.56. This implies a huge improvement of 38\% in terms of F1-Score. Besides this, our least performing transformer model, DistilBERT, offers an F1-Score of 0.74, which is also 32\% higher than that of SVM.

\item \textbf{\textit{Usability}} RoBERTa shows the best performance with a Precision of 0.68, Recall of 0.69, and an F1-Score of 0.68 among other deep and shallow models. Unlink the previous aspect (Performance), RoBERTa improves the baseline by only 10\% in terms of F1-Score. Despite the fact that transformers outperform the baseline SVM, the F1-Score of XLNet and SVM is almost the same, i.e., 0.63 for XLNet and 0.62 for SVM.

\item \textbf{\textit{Security}} Unlike the previous two aspects, DistilBERT shows the best F1-Score of 0.76 for security aspect detection, although RoBERTa has the best Precision of 0.81 and BERT has the highest Recall of 0.81. Our best performing model for security aspects, DistilBERT, increases the baseline SVM by 27\% in terms of F1-Score. Other transformer models also show an almost similar improvement over the baseline SVM.

\item \textbf{\textit{Community}} We found an interesting result for this aspect, as the expected transformer models fail to cross the performance line of basline SVM. In fact, the best performing transformer model, DistilBERT, has a massive gap of 30\% between the baseline SVM. In addition, we found that BERT records the lowest F1-Score of 0.13 among the transformer models.

\item \textbf{\textit{Compatibility}} The transformer models perform poorly in detecting this aspect. At best, the DistilBERT achieves a maximum F1-Score of 0.13, which is exactly the same as the baseline SVM. Nevertheless, all the transformers and baseline SVM suffer from lower recall. Both BERT and XLNet record the lowest recall of 0.01.

\item \textbf{\textit{Portability}} We observed that the baseline SVM narrowly beats the transformer models with an F1-Score of 0.61. Although RoBERTa offers better Precision (0.67) than SVM, the model unfortunately suffers from low recall. We also found that the XLNet model has the lowest precision (0.33) and recall (0.19) among all models for this task.

\item \textbf{\textit{Documentation}} RoBERTa is the leading performer when it comes to detecting this aspect. It has the highest Precision of 0.69 and Recall of 0.57, which results in the best F1-Score of 0.61. It comprehensively outperforms the baseline SVM model by 25\% in terms of F1-Score. Other transformers also shows similar results, but lower than the RoBERTa.

\item \textbf{\textit{Bug}} We observed that both BERT and DistilBERT have the exact same F1-Score value of 0.63, which is also the highest among all models. However, both models have different Precision and Recall. DistilBERT has better Recall (0.66) whereas BERT has higher Precision (0.70). These models also outperform the baseline SVM by a long distance as well.

\item \textbf{\textit{Legal}} We again found that DistilBERT, having an F1-Score of 0.63, appears to be superior over other transformers for this aspect detection. Although it has only a 3\% higher F1-Score than RoBERTa, it thoroughly tops BERT's F1-Score of 0.38 and XLNet's F1-Score of 0.35. In addition, DistilBERT has better Precision, Recall, and F1-Score as well.

\item \textbf{\textit{Only Sentiment}} For this aspect, RoBERTa offers the best F1-Score of 0.58, whereas BERT shows the lowest F1-Score of 0.54 among the transformer models. Nevertheless, all these models are able to beat the benchmark performance of the SVM model.

\item \textbf{\textit{Others}} All the transformer models perform similarly when detecting these aspects. The best performing model, RoBERTa, has only a 5\% higher F1-Score than the lowest performing model, XLNet. However, these models improves the F1-Score of the baseline SVM by more than 8\%.

\end{enumerate}
We therefore found RoBERTa as the best performer for Performance, Usability, Documentation, OnlySentiment, and Others aspects detection and DistilBERT for Security, Bug, Compatibility, and Legal aspects detection. This also implies that XLNet and BERT are not as useful as RoBERTa and DistilBERT for software aspect detection. In addition, no transformer models are as good as baseline SVM for detecting Portability and Community aspects.

\subsection{Hyper-parameter settings}
\label{sec6:hyperparameter}

The hyper-parameters associated with different models catalyze the performance of the model by a lot. Choosing the correct value of hyper-parameters for the best performance is not as easy task. In the conducted experiment many variation of hyper-parameters were tested in order to get good results in the experiment. Some of these values can be seen in Table ~\ref{tab:perf_best}. The other experiments have been conducted with different combinations quite similar to the ones seen in Table ~\ref{tab:perf_best}. 

\begin{table}[h]
    \centering
    \caption{Best performing models by precision with hyper-parameters}
    \label{tab:perf_best}
    \resizebox{!}{!}{\begin{tabular}{|c|c|c|c|c|}
    \hline
    \textbf{Metric} & \textbf{Best Model} & \textbf{Batch Size} & \textbf{Epoch} & \textbf{Learning Rate} \\ \hline
    Performance &   RoBERTa &   32  &   3   &   1.00E-05 \\
    Usability   &   RoBERTa &   32  &   3   &   1.00E-05 \\
    Security    &   DistilBERT  &   16  &   2   &   1.00E-05 \\
    Community   &   DistilBERT  &   16  &   3   &   2.00E-05 \\
    Compatibility   &   DistilBERT  &   16  &   3   &   2.00E-05 \\
    Portability &   DistilBERT  &   16  &   3   &   2.00E-05 \\
    Documentation   &   RoBERTa &   32  &   2   &   1.00E-05 \\
    Bug &   BERT    &   32  &   3   &   3.00E-05 \\
    Legal   &   DistilBERT &   32  &   3   &   1.00E-05 \\
    OnlySentiment   &   RoBERTa &   32  &   3   &   1.00E-05 \\ \hline
    \end{tabular}}
\end{table}

\subsection{Result Analysis}
We described our experimental results in Section \ref{sec: experimental_result}. We discussed the performance of each individual models and also showed a comparative analysis among them. We further investigated the results and found some interesting findings. In this section, we are focusing on these findings. We show a detailed analysis of our findings as follows:
\begin{enumerate}
    \item \textbf{\textit{Transformer Models are more effective than baseline shallow models, such as, SVM and Logits.}} Previously, Uddin at el. \cite{uddin2017automatic} showed that a shallow model can apprehend software aspects. However, our approach to detect software aspects is more effective. Our transformer models improve the performance of most of the aspects. Our best performing model for each aspect has higher precision and recall than baseline SVM. In addition, we find that our approach has a better balance between precision and recall, which indicates the stability of the model. For example, RoBERTa achieves a recall of 0.74 and a precision of 0.80 for `Performance' aspect, which is around only 4.5\% deviation from the F1 score. However, baseline SVM has a 21.8\% deviation between F1-score and precision and recall. This indicates that our optimized transformer model is more effective and reliable than previous studies on aspects detection. 
    \item \textbf{\textit{Machine learning tools fail to apprehend `Community' and `Compatibility' aspects.}} Although transformer models do a fairly good job of detecting software aspects, they completely go wrong when it comes to detecting `Community' or `Compatibility' aspects. An optimized model like RoBERTa has only 18\% and 17\% precision for Community and Compatibility aspect detection, respectively. One of the possible reasons for such low performance can be attributed to the low positive rates of these aspects in the dataset. However, the baseline model, SVM, which requires a lower training sample, shows similar performance. This indicates that these two aspects are difficult to comprehend for ML tools. This is because there are implicit contexts sparsely distributed in the dataset, which makes the classification task more challenging compared to other aspects. An extended dataset with more positive samples could be a fair attempt to improve the benchmark, which we left as our future work.
    \item \textbf{\textit{Performance of transformer models varies depending on the aspects.}} Transformer models perform fairly well despite a low positive rate of some software aspects, such as, `Legal'. However, the performance changes when the targeted aspects have been altered, even for the same model. For example, RoBERTa does pretty well for `Performance' aspect, but for `Security' aspect, the performance drops a bit. A closer look at the results also indicates that there are no models that perform well across all aspects detection among the studied models. For instance, RoBERTa performs well for the usability aspect, whereas DistilBERT performs better for `Security' aspects. This finding urges immediate research into multi-class aspects detection by ensembling all these models. In the future, we can explore this to learn the effectiveness of such blending for this aspect of the detection task.
    \item \textbf{\textit{Larger architectures like XLNet are ineffective in interpreting software aspects compared to smaller architectures like DistilBERT.}} An interesting finding of our studies is that larger or unoptimized architectural models perform lower than smaller or optimized architectural models. According to our background studies, BERT and XLNet have the largest architectures, where DistilBERT is a distiled version of BERT and RoBERTa is the most optimized model. Our experimental results imply that both BERT and XLNet models have lower performance metrics than RoBERTa and DistilBERT. Even though DistilBERT shares the same architecture, the performance shows huge improvement over BERT for all aspects. This finding could be attributed to the lower sample size for most of the aspects. However, for `Others' aspect, which has enough samples, DistilBERT and RoBERTa also outclass BERT and XLNet. This result notes that huge parameter lists of larger architecture cause unnecessary intricacies in predicting software aspects, which eventually result in a partial drop in performance metrics. We believe this requires further investigation to identify the more literal cause of such behaviors. We left this work as a future avenue of improvements.    
\end{enumerate}

\section{Conclusion and Future Work}
\label{sec:conclusion}

The conducted experiments looked into how different transformer models performed on a benchmark dataset that had already been used with the SVM model. When it comes to predicting aspects of a text involving an API, the study found that different transformer architectures may outperform SVM. Despite the fact that there is no one architecture that can accurately categorize the text contexts of all sorts of aspects, the experiments yielded several ideas. The greatest issue that surfaced was the restriction of unbalanced data. Furthermore, the findings suggest that models with complicated large architecture have a decreased possibility of producing effective classification results in this experiment. By looking at the performance score, it's evident that there's still room for progress in this field of study. Though the study in this paper focuses on employing several types of transformer models, it does not evaluate all of the models available. There are many other sorts of models that may be employed to construct a system with higher prediction skills in the future. Furthermore, the dataset is rather unbalanced, and balancing it might considerably enhance the findings. Overall, because the necessity of precise API selection is so critical, subsequent research based on the results of the experiments has a lot of promise. 
%
%
%

\begin{thebibliography}{8}
\bibitem{cc-intro}
Creative Commons license - Wikipedia, \url{https://en.wikipedia.org/wiki/Creative\_Commons\_license}, accessed: 01 Apr, 2022

\bibitem{stackoverflow-wiki}
Stack Overflow - Wikipedia, \url{https://en.wikipedia.org/wiki/Stack_Overflow}, ac-
cessed: 04 Apr, 2022

\bibitem{transformer}
Vaswani, Ashish, Noam Shazeer, Niki Parmar, Jakob Uszkoreit, Llion Jones, Aidan N. Gomez, Łukasz Kaiser, and Illia Polosukhin. "Attention is all you need." \textsl{Advances in neural information processing systems} 30 (2017).

\bibitem{bert}
Devlin, Jacob, Ming-Wei Chang, Kenton Lee, and Kristina Toutanova. "Bert: Pre-training of deep bidirectional transformers for language understanding." \textsl{arXiv preprint arXiv:1810.04805} (2018).

\bibitem{roberta}
Liu, Yinhan, Myle Ott, Naman Goyal, Jingfei Du, Mandar Joshi, Danqi Chen, Omer Levy, Mike Lewis, Luke Zettlemoyer, and Veselin Stoyanov. "Roberta: A robustly optimized bert pretraining approach." \textsl{arXiv preprint arXiv:1907.11692} (2019).

\bibitem{distilbert}
Sanh, Victor, Lysandre Debut, Julien Chaumond, and Thomas Wolf. "DistilBERT, a distilled version of BERT: smaller, faster, cheaper and lighter." \textsl{arXiv preprint arXiv:1910.01108} (2019).

\bibitem{xlnet}
Yang, Zhilin, Zihang Dai, Yiming Yang, Jaime Carbonell, Russ R. Salakhutdinov, and Quoc V. Le. "Xlnet: Generalized autoregressive pretraining for language understanding." \textsl{Advances in neural information processing systems} 32 (2019).

\bibitem{barua}
A. Barua, S. W. Thomas, and A. E. Hassan. "What are developers talking about? an analysis of topics and trends in stack overflow." \textsl{Empirical Software Engineering}, pages 1-31, 2012.

\bibitem{gias_emp}
Nibir Mandal and Gias Uddin. "An empirical study of IoT security aspects at sentence-level in developer textual discussions." \textsl{Information and Software Technology} 150, Pages- 106970 (2022).

\bibitem{blei}
D. M. Blei, A. Y. Ng, and M. I. Jordan. "Latent dirichlet allocation." \textsl{Journal of Machine Learning Research}, 3(4-5):993-1022, 2003.

\bibitem{kavaler}
D. Kavaler, D. Posnett, C. Gibler, H. Chen, P. Devanbu, and V. Filkov. "Using and asking: Apis used in the android market and asked about in stackoverflow." \textsl{In Proceedings of the International Conference on Social Informatics}, pages 405–418, 2013.

\bibitem{parnin}
C. Parnin, C. Treude, L. Grammel, and M.-A. Storey. "Crowd documentation: Exploring the coverage and dynamics of api discussions on stack overflow." \textsl{Technical report, Technical Report GIT-CS-12-05}, Georgia Tech, 2012.

\bibitem{gias_auto}
Uddin, Gias, and Foutse Khomh. "Automatic mining of opinions expressed about apis in stack overflow." \textsl{IEEE Transactions on Software Engineering} 47, no. 3 (2019): 522-559.

\bibitem{gias_mining}
Gias Uddin and Foutse Khomh. "Mining API aspects in API reviews." \textsl{Technical report}, 2017.

\bibitem{gias_summary}
Gias Uddin and Foutse Khomh. "Opiner: an opinion search and summarization engine for apis." \textsl{In Proceedings of the $32^nd$ IEEE/ACM International Conference on Automated Software Engineering (ASE 2017)}, pages 978–983. IEEE Computer Society, 2017.

\bibitem{uddin2017automatic}
Uddin, Gias, and Foutse Khomh. "Automatic summarization of API reviews." \textsl{In 2017 $32^nd$ IEEE/ACM International Conference on Automated Software Engineering (ASE)}, pp. 159-170. IEEE, 2017.

\bibitem{uddin2019understanding}
Uddin, Gias, Olga Baysal, Latifa Guerrouj, and Foutse Khomh. "Understanding how and why developers seek and analyze API-related opinions." \textsl{IEEE Transactions on Software Engineering} 47, no. 4 (2019): 694-735.

\bibitem{lin_bin}
Lin, Bin, Fiorella Zampetti, Gabriele Bavota, Massimiliano Di Penta, and Michele Lanza. "Pattern-based mining of opinions in Q\&A websites." \textsl{In 2019 IEEE/ACM $41^st$ International Conference on Software Engineering (ICSE)}, pp. 548-559. IEEE, 2019.

\bibitem{liu}
Minqing Hu and Bing Liu. "Mining and summarizing customer reviews." \textsl{In Proceedings of the $10^th$ ACM SIGKDD International Conference on Knowledge Discovery and Data Mining (SIGKDD 2004)}, pages 168–177. ACM, 2004.

\bibitem{shaohua}
Shaohua Wang, NhatHai Phan, Yan Wang, and Yong Zhao. "Extracting API tips from developer question and answer websites." \textsl{In Proceedings of the $16^th$ International Conference on Mining Software Repositories (MSR 2019)}, pages 321–332. IEEE / ACM, 2019.

\bibitem{zhang}
Yingying Zhang and Daqing Hou. "Extracting problematic API features from forum discussions." \textsl{In Proceedings of the IEEE $21^st$ International Conference on Program Comprehension (ICPC 2013)}, pages 142–151. IEEE Computer Society, 2013.

\bibitem{example-api}
Top 50 Most Popular APIs (Updated for 2022) | RapidAPI, \url{https://rapidapi.com/blog/most-popular-api/}, accessed: 11 Apr, 2022

\bibitem{transferlearning}
Raffel, Colin, Noam Shazeer, Adam Roberts, Katherine Lee, Sharan Narang, Michael Matena, Yanqi Zhou, Wei Li, and Peter J. Liu. "Exploring the limits of transfer learning with a unified text-to-text transformer." \textsl{arXiv preprint arXiv:1910.10683} (2019).

\end{thebibliography}
%

\end{document}